# Resource Allocation in 6G Optical Wireless Systems

Osama Zwaid Alsulami, T. E. H. El-Gorashi and Jaafar M. H. Elmirghani

School of Electronic and Electrical Engineering, University of Leeds, LS2 9JT, United Kingdom

ml15ozma@leeds.ac.uk, t.e.h.elgorashi@leeds.ac.uk, j.m.h.elmirghani@leeds.ac.uk

## Abstract

The abundant optical spectrum is a promising part of the electromagnetic spectrum for 6G communication systems. The visible light spectrum which is a part of the optical spectrum, can be used to provide communication and illumination simultaneously. Visible light communication (VLC) systems have been widely researched, however, little work has focused on the area of multiple access. This chapter studies wavelength division multiple access (WDMA) techniques in VLC systems to support multiple users. In addition, the optimization of resource allocation is considered in this chapter by developing a mixed-integer linear programming (MILP) model that can be used to maximize the signal to noise and interference ratio (SINR) while supporting multiple users. The optimized resource allocation results in the best assignment of access points (APs) and wavelengths to users. Different indoor environments such as office, data center and aircraft cabins are evaluated in this chapter. A laser diode (LD) with four wavelengths (red, green, yellow and blue) is used to provide high bandwidth for communication and white light for illumination. Also, an angle diversity receiver (ADR) is utilized to receive signals and reduce noise and interference by exploiting the spatial domain.

## 1 Introduction

The number of users connected to the Internet has increased significantly and there is a growing demand for high data rates. The radio spectrum, which is the current spectrum that is broadly used in the indoor environment, faces various limitations, such as it being a scarce and congested spectrum resulting in limited channel capacity and low transmission rates. Thus, several techniques have been introduced to minimize such limitations in the spectrum, including advanced modulation, smart antennas and the concept of multiple-input and multiple-output (MIMO) systems [1], [2]. Achieving high data rates beyond 10 Gbps for each user is challenging when using the current radio spectrum, especially if the number of connected devices continues to increase. Cisco expects the increase in the number of connected devices from 2017 to 2021 to be approximately 27 times. Therefore, an alternative spectrum should be discovered to meet these demands. The optical spectrum is one such spectrum that promises to support multiple users at high data rates. Optical wireless communication (OWC) systems have been introduced by researchers as a potential solution. OW systems have important advantages, such as excellent channel characteristics, abundant bandwidth and low cost components compared to the current radio frequency (RF) wireless technology [3] – [10]. OWC systems can provide Tb/s aggregate data rates in an indoor environment for the sixth generation wireless communication. OWC data rates exceeding 25 Gbps per user have been demonstrated in downlink communication in indoor settings [9] – [20]. The uplink has been studied in different works [21], [22], also, energy efficiency is an area that must be given an attention [23]. In addition, various configurations of transmitter and receiver have been studied to reduce the delay spread [15], [24] – [31]while increasing the signal-to-noise ratio (SNR) [32] – [36]. Different multiple access methods used in radio frequency (RF) systems, such as sharing the time, frequency, wavelength or code domains, can be used in OWC systems to support multiple users. Utilizing resources efficiently is required to avoid signal quality degradation. Thus, attention has been given to methods that can be used to efficiently share the OWC resources including space, power, wavelength and time resources.

This chapter introduces an indoor OWC system that provides multiple access using wavelength division multiple access (WDMA). The optimization of resource allocation in terms wavelengths and access points to different users has been consider in this work by maximizing the sum over all users of Signal to Interference-plus-Noise Ratio (SINRs). A Mixed Integer Linear Programme (MILP) model has been developed to optimize the resource allocation. The rest of this chapter is organized as follows: Section 2 describes the transmitter and receiver design, Section 3 outlines the relevant multiple access techniques including WDMA and the developed MILP model. The simulation setup and results in different indoor environments are given in Section 4 and finally Section 5 presents the conclusions.

## 2 Transmitter and Receiver Design

### 2.1 Transmitter Design

In this work a visible light illumination source that utilizes red, yellow, green and blue (RYGB) laser diodes (LDs) is also used for as a transmitter (access point (AP)) for communication at the same time. RYGB LDs are used to provide four wavelengths, as shown in Fig. 1a, and each wavelength can carry a different data stream from to support multiple users. The LDs are used to offer high bandwidth to transfer data and support high data rates. RYGB LDs can be used safely in indoor environments to provide illumination, as stated in [37]. Four mixed LDs colours can be used to offer white illumination, as shown in Fig. 1b. The white colour source is generated by combining the four different LD colours utilizing chromatic beam-splitters. The combined beams are reflected using a mirror and subsequently pass through a diffuser to reduce speckle before the illumination is passed to the room environment.

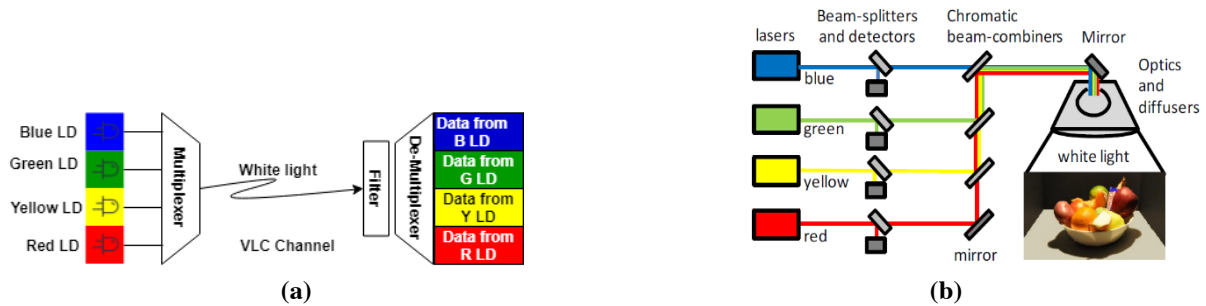

**Fig. 1** WDM used in VLC systems using LDs.

## 2.2 Receiver Design

In this work, an angle diversity receiver (ADR) that consists of four branches (See Fig. 3) was used in the three different indoor environments to collect signals from different directions. Each of the four branches of the ADR has a narrow field of view (FOV) to reduce interference from other users and inter-symbol interference (ISI). The design of the ADR is similar to [9], [22], [38]. Each branch of the ADR covers different areas in the indoor environment by orienting the branch to that area based on Azimuth ($Az$) and Elevation ($El$) angles. These angles were set as follow: The $Az$ angles of the four branches were set to be 45°, 135°, 225° and 315°, whereas, the $El$ angles of these branches were set to 70°. The FOV of each branch of the receiver has been chosen to be 21° as a narrow FOV reduces both interference and ISI. Each branch has a detector area of 10 $mm^2$ and responsivity equal to 0.4, 0.35, 0.3 and 0.2 A/W for red, yellow, green and blue wavelengths respectively.

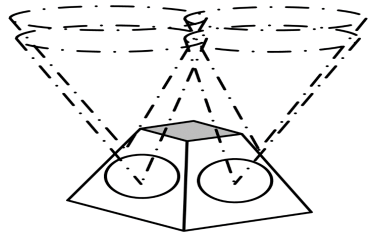

**Fig. 2** Angle Diversity Receiver (ADR).

# 3 Multiple access

### 3.1 Wavelength Division Multiple Access (WDMA)

Several Multiple access (MA) techniques that are utilized in RF systems can also be utilized in OWC systems. Wavelength division multiple access (WDMA) has been a subject of interest in many studies in OWC systems to support multiple users by sharing wavelengths among users [6], [38] – [41]. A multiplexer is used at the transmitter to aggregate wavelengths into a single optical beam. The receiver uses a de-multiplexer to separate the wavelengths. Two different light sources are utilized in OWC systems, namely RGB LEDs and RYGB LDs and both can support WDMA (see Fig. 4) [6], [42].

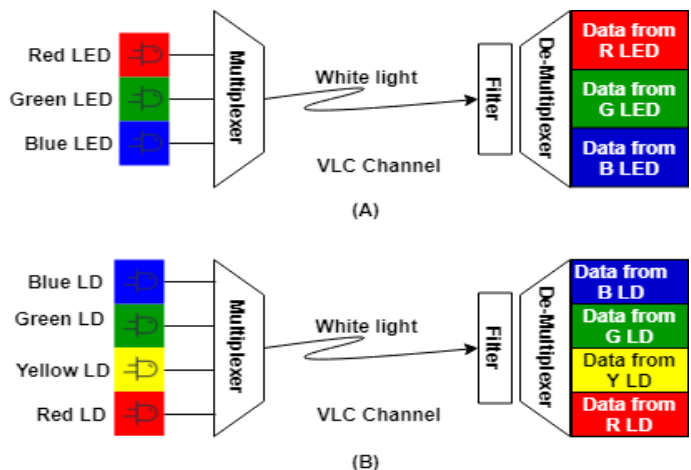

**Fig. 3** WDM used in VLC systems (A) LDs transmitter and (B) WDM receivers.

## 3.2 MILP model

This section presents a MILP model developed to optimize resource allocation based on maximizing the sum of SINRs for all possible users [38]. This MILP model is used to assign users or devices to access points (APs) and wavelengths while maximizing the sum of SINRs over all users. The input data of the MILP model is the pre-calculated received power, receiver noise and background noise which can then be used to measure all users' SINRs in each location of interest in the indoor environment. A simple example to show how the assignment is done is provided in Fig. 4, where three users, three APs and two wavelengths (Red and Blue) are used. User 1 is the only one assigned to the blue wavelength and only suffers from the background noise of the other APs. However, Users 2 and 3 share the same wavelength (Red) from different APs and they thus interfere with one

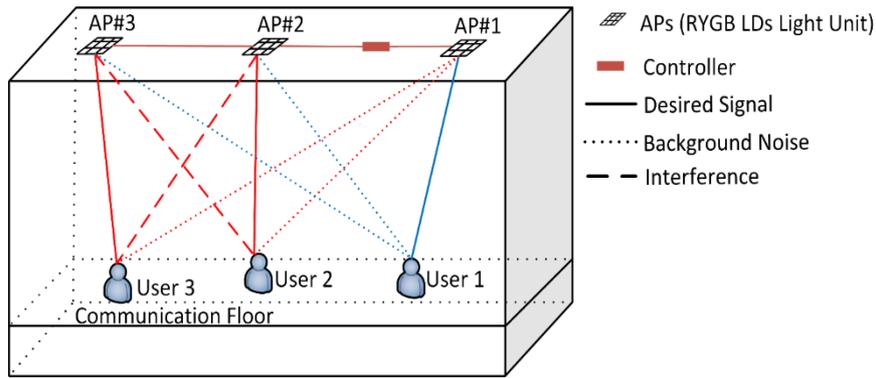

another and receive background noise from AP 1.

**Fig. 4** Example of assignements in the MILP model in an indoor OWC environment.

The developed MILP model consists of sets, parameters and variables that are used to obtain the assignments. Tables 1, 2 and 3 introduce these sets, parameters and variables respectively.

**Table 1** Sets of the MILP model

| | |
|---|---|
| $\mathcal{US}$ | Set of users in the indoor environment; |
| $\mathcal{AP}$ | Set of access points; |
| $\mathcal{W}$ | Set of wavelengths; |
| $\mathcal{B}$ | Set of receiver branches. |

**Table 2** Parameters of the MILP model

| | |
|---|---|
| $us, ui$ | $us$ is desired user, $ui$ refers to other users; |
| $ap, cp$ | $ap$ is the allocated access point to user $us$, $cp$ is the allocated access point to another users $ui$; |
| $\lambda$ | Refers to wavelength; |
| $b, f$ | Receiver branch, for the desired user $us$ and the other users $ui$; |
| $P_{us,b}^{ap,\lambda}$ | The electrical received power at the receiver from user $us$ due to the received optical power from the access point $ap$ using the wavelength $\lambda$ at receiver branch $b$; |
| $\sigma_{us,b}^{cp,\lambda}$ | The shot noise at the receiver of user $us$ due to the background unmodulated power from access point $cp$ using wavelength $\lambda$ at receiver branch $b$; |
| $\sigma_{Rx}$ | The receiver noise; |

**Table 3** Variables of the MILP model

| | |
|---|---|
| $USINR_{us,b}^{ap,\lambda}$ | SINR of user *us* allocated to access point *ap* using wavelength $\lambda$ at receiver branch *b*; |
| $S_{us,b}^{ap,\lambda}$ | A binary selector function where 1 refers to the assignment of user *us* to access point *ap* using wavelength $\lambda$ at receiver branch *b*; |
| $\phi_{us,ui,b,f}^{ap,cp,\lambda}$ | Linearization variable that is not negative value, $\phi_{us,ui,b,f}^{ap,cp,\lambda} = USINR_{us,b}^{ap,\lambda} S_{ui,f}^{cp,\lambda}$; |

The Objective of the MILP model is to maximize the sum of all users' SINRs which can be defined as follows:

$$Maximise \sum_{us \in US} \sum_{ap \in AP} \sum_{\lambda \in W} \sum_{b \in B} USINR_{us,b}^{ap,\lambda}, \quad (1)$$

The SINR of each user can be calculated as follows:

$$USINR_{us,b}^{ap,\lambda} = \frac{Signal}{Interference + Noise}$$

$$= \frac{P_{us,b}^{ap,\lambda} S_{us,b}^{ap,\lambda}}{\sum_{\substack{cp \in AP \\ cp \neq ap}} \sum_{\substack{ui \in US \\ ui \neq us}} \sum_{f \in B} P_{us,b}^{cp,\lambda} S_{ui,f}^{cp,\lambda} + \sum_{\substack{cp \in AP \\ cp \neq ap}} \sigma_{us,b}^{cp,\lambda} \left[1 - \sum_{\substack{ui \in US \\ ui \neq us}} \sum_{f \in B} S_{ui,f}^{cp,\lambda}\right] + \sigma_{Rx}}, \quad (2)$$

where the receiver noise $\sigma_{Rx}$ can be given by:

$$\sigma_{Rx} = N_R B_E, \quad (3)$$

where $N_R$ is the density of receiver noise measured in ($A^2$/Hz) and $B_E$ is the electrical bandwidth.

The background noise $\sigma_{us,b}^{cp,\lambda}$ can be evaluated as:

$$\sigma_{us,b}^{cp,\lambda} = 2e(R\ PO_{us,b}^{cp,\lambda})B_o B_e, \quad (4)$$

where $R$ is the receiver responsivity, $e$ is the electron charge, $PO_{us,b}^{cp,\lambda}$ is the received optical power from unmodulated APs and $B_o$ is the optical bandwidth.

The electrical received power $P_{us,b}^{ap,\lambda}$ can be calculated as:

$$P_{us,b}^{ap,\lambda} = (R\ PO_{us,b}^{ap,\lambda})^2, \quad (5)$$

where $PO_{us,b}^{ap,\lambda}$ is the receiver optical power.

The numerator in Equation 1 is the received signal multiplied by the selector binary variable; while, the denominator consists of two parts: The first part is the sum of interferences coming from all APs that use the same wavelength for communication, and the second part is the sum of the background noise coming from all unmodulated APs using the same wavelength.

Rewriting Equation (1) gives:

$$\sum_{\substack{cp \in AP \\ cp \neq ap}} \sum_{\substack{ui \in US \\ ui \neq us}} \sum_{f \in B} USINR_{us,b}^{ap,\lambda} P_{us,b}^{cp,\lambda} S_{ui,f}^{cp,\lambda} + \sum_{\substack{cp \in AP \\ cp \neq ap}} USINR_{us,b}^{ap,\lambda} \sigma_{us,b}^{cp,\lambda} \left[1 - \sum_{\substack{ui \in US \\ ui \neq us}} \sum_{f \in B} S_{ui,f}^{cp,\lambda}\right] + USINR_{us,b}^{ap,\lambda} \sigma_{Rx}$$

$$= P_{us,b}^{ap,\lambda} S_{us,b}^{ap,\lambda}, \quad (6)$$

Equation (6) can be rewritten as:

$$\sum_{\substack{cp \in AP \\ cp \neq ap}} \sum_{\substack{ui \in US \\ ui \neq us}} \sum_{f \in B} (P_{us,b}^{cp,\lambda} - \sigma_{u,f}^{b,\lambda})\ USINR_{us,b}^{ap,\lambda} S_{ui,f}^{cp,\lambda} + \sum_{\substack{cp \in AP \\ cp \neq ap}} USINR_{us,b}^{ap,\lambda} \sigma_{us,b}^{cp,\lambda} + USINR_{us,b}^{ap,\lambda} \sigma_{Rx} = P_{us,b}^{ap,\lambda} S_{us,b}^{ap,\lambda}, \quad (7)$$

The first part of Equation (6) consists of interference and background noise which is a nonlinear part composed of the multiplication of a continuous variable and a binary variable. To linearize this part, the same equations as [38], [43] were utilized which introduces a linearization variable that is non-negative; $\phi_{us,ui,b,f}^{ap,cp,\lambda} = \gamma_{us,b}^{ap,\lambda} S_{ui,f}^{cp,\lambda}$ as shown below:

$$\phi_{us,ui,b,f}^{ap,cp,\lambda} \geq 0. \tag{8}$$

$$\phi_{us,ui,b,f}^{ap,cp,\lambda} \leq \alpha S_{ui,f}^{cp,\lambda}, \quad \forall us, ui \in \mathcal{US}, \forall ap, cp \in \mathcal{AP}, \forall \lambda \in \mathcal{W}, \forall b, f \in \mathcal{B} \quad (us \neq ui, ap \neq cp) \tag{9}$$

where $\alpha$ is a very large value, $\alpha \gg USINR$.

$$\phi_{us,ui,b,f}^{ap,cp,\lambda} \leq USINR_{us,b}^{ap,\lambda}, \quad \forall us, ui \in \mathcal{US}, \forall ap, cp \in \mathcal{AP}, \forall \lambda \in \mathcal{W}, \forall b, f \in \mathcal{B} \quad (us \neq ui, ap \neq cp) \tag{10}$$

$$\phi_{us,ui,b,f}^{ap,cp,\lambda} \geq \alpha S_{ui,f}^{cp,\lambda} + USINR_{us,b}^{ap,\lambda} - \alpha, \quad \forall u, m \in \mathcal{U}, \forall a, b \in \mathcal{A}, \forall \lambda \in \mathcal{W}, \forall f \in \mathcal{B} \quad (u \neq m, a \neq b) \tag{11}$$

Using linearization Equations (8-11) to replace the first part of Equation (7), Equation (7) can be re-written as follows:

$$\sum_{\substack{cp \in \mathcal{AP} \\ cp \neq ap}} \sum_{\substack{ui \in \mathcal{US} \\ ui \neq us}} \sum_{f \in \mathcal{B}} (P_{us,b}^{cp,\lambda} - \sigma_{u,f}^{b,\lambda}) \phi_{us,ui,b,f}^{ap,cp,\lambda} + \sum_{\substack{cp \in \mathcal{AP} \\ cp \neq ap}} USINR_{us,b}^{ap,\lambda} \sigma_{us,b}^{cp,\lambda} + USINR_{us,b}^{ap,\lambda} \sigma_{Rx} = P_{us,b}^{ap,\lambda} S_{us,b}^{ap,\lambda}, \tag{12}$$

The MILP model is subject to three constraints as follows:

The first constraint ensures that a wavelength that belongs to an AP can only be allocated once and can be expressed as follows:

$$\sum_{us \in \mathcal{US}} \sum_{b \in \mathcal{B}} S_{us,b}^{ap,\lambda} \leq 1, \quad \forall ap \in \mathcal{AP}, \forall \lambda \in \mathcal{W} \tag{13}$$

The second constraint ensures that all users are assigned to one access point, one wavelength and one branch (due to the use of selection combining (SC) in the ADR receiver), i.e., one receiver branch is selected per user and can be written as follows:

$$\sum_{ap \in \mathcal{AP}} \sum_{\lambda \in \mathcal{W}} \sum_{b \in \mathcal{B}} S_{us,b}^{ap,\lambda} = 1, \quad \forall us \in \mathcal{US} \tag{14}$$

The last constraint ensures that the SINR of each user should not go below 15.6 dB, which is the threshold, in order to provide a bit error rate (BER) $10^{-9}$ using on-off keying (OOK) modulation. This constraint can be expressed as follows:

$$USINR_{us,b}^{ap,\lambda} \geq 10^{\frac{36}{10}}, \quad \forall us \in \mathcal{US}, \forall ap \in \mathcal{AP}, \forall \lambda \in \mathcal{W}, \forall b \in \mathcal{B} \tag{15}$$

The CPLEX solver was used to solve the MILP model.

# 4 Evaluation in different indoor environments

Three different indoor environments were evaluated in this work; office, cabin and data centre. The optical channel is modelled using a ray tracing algorithm similar to [44] and [45]. In addition, up to the second order reflection were modelled in this work due to the third and higher order reflections having a very low impact on the received optical power [44]. The office was divided into surfaces and one of these surfaces was divided into many small elements. These elements reflect signals following a Lambertian model and act as the secondary small emitters [46]. The area of the elements has a significant impact on the resolution of the results. Increasing the element's area, results in a reduction in the temporal resolution of the results. However, the simulation running time increases when the element's area is reduced.

### 4.1 Office

An empty office was examined in this work. The following sections introduce the system configuration, evaluation setup and results in a multi-user scenario.

#### 4.1.1 Office OWC System Configuration

In this section, an unfurnished office without doors and windows is considered, as shown in Fig. 5. Table 4 introduces the OWC system parameters used.

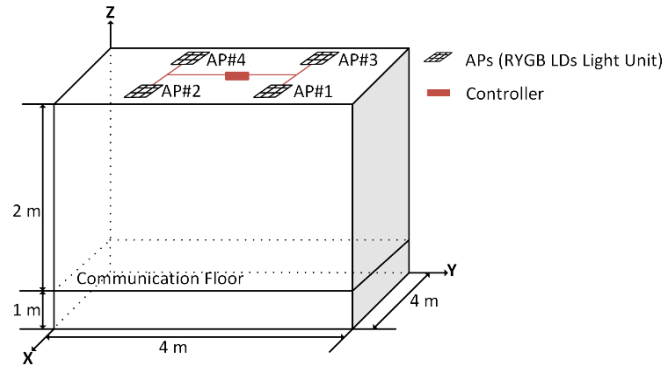

**Fig. 5** Office dimensions.

**Table 4** The Office OWC Parameters

| Parameters | Configurations | |
|---|---|---|
| **Office** | | |
| Walls and ceiling reflection coefficient | 0.8 [51] | |
| Floor reflection coefficient | 0.3 [51] | |
| Number of reflections | 1 | 2 |
| Area of reflection element | 5 cm × 5 cm | 20 cm × 20 cm |
| Order of Lambertian pattern, walls, floor and ceiling | 1 [51] | |
| Semi-angle of reflection element at half power | 60° | |
| **Transmitters** | | |
| Number of transmitters' units | 4 | |
| Transmitters locations (x, y, z) | (1 m, 1 m, 3 m), (1 m, 3 m, 3 m), (3 m, 1 m, 3 m) and (3 m, 3 m, 3 m) | |
| Number of RYGB LDs per unit | 9 | |
| Transmitted optical power of Red, Yellow, Green and Blue LD | 0.8, 0.5, 0.3 and 0.3 W | |
| Semi-angle at half power | 60° | |
| **Receiver** | | |
| Receiver noise current spectral density | 4.47 pA/√Hz [52] | |
| Receiver bandwidth | 5 GHz | |

### 4.1.2 Office OWC System Setup and Results

A scenario of 8 users was examined in this part where two users were placed under each AP. Table 5 shows the locations of the 8 users and the optimized resource allocation using the MILP model described in Section 3.2.

**Table 5** The optimized resource allocation of APs and wavelengths.

| User | Location (x, y, z) | Access Point | Wavelength | Receiver Branch |
|---|---|---|---|---|
| 1 | (0.5m, 0.5m, 1m) | 1 | Yellow | 1 |
| 2 | (0.5m, 2.5m, 1m) | 2 | Red | 1 |
| 3 | (1.5m, 1.5m, 1m) | 1 | Red | 3 |
| 4 | (1.5m, 3.5m, 1m) | 2 | Yellow | 3 |
| 5 | (2.5m, 0.5m, 1m) | 3 | Red | 1 |
| 6 | (2.5m, 2.5m, 1m) | 4 | Yellow | 1 |
| 7 | (3.5m, 1.5m, 1m) | 3 | Yellow | 3 |
| 8 | (3.5m, 3.5m, 1m) | 4 | Red | 3 |

Fig. 6 shows the channel bandwidth, SINR and data rate of the 8 user scenario. All user locations support a high channel bandwidth above 8 GHz. In addition, all user locations can provide high SINR above the threshold (15.6 dB). However, users assigned to the Yellow wavelength have a lower SINR compared to those assigned to Red wavelength. The reason for that is that the Yellow wavelength has a lower transmitted power (needed to ensure the version of white colour desired [38]) and receiver responsivity compared to the Red wavelength. In terms of data rate, all users can support a data rate above 7 Gbps.

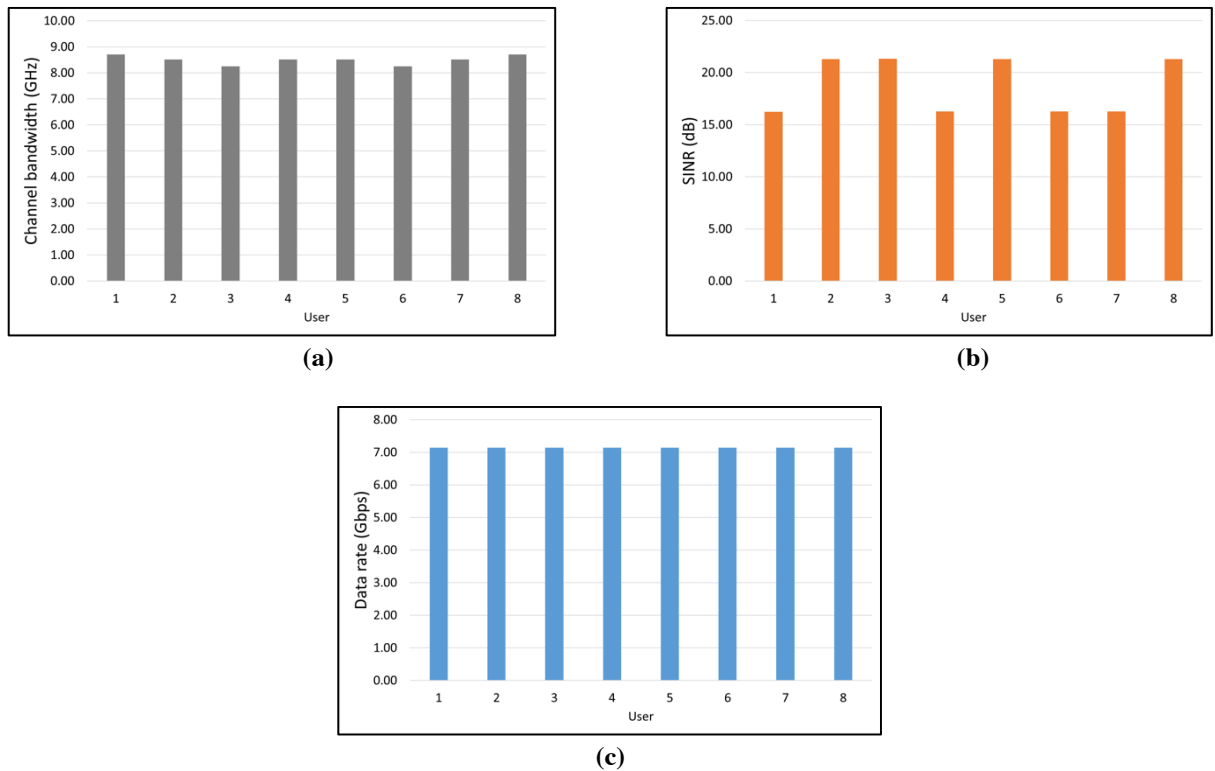

**Fig. 6 (a)** Channel Bandwidth, **(b)** SINR and **(c)** Data Rate.

### 4.2 Aircraft Cabin

An aircraft cabin downlink is evaluated in this section to support multiple users. OWC does not contribute further electromagnetic interference and is therefore attractive in this environment. The system configuration, including the type of aircraft, is shown in the next section. The scenario evaluated and the result are provided in the following section.

### 4.2.1 Aircraft Cabin OWC System Configuration

The type of aircraft examined as an indoor environment in this work is an Airbus A321neo (see Fig. 7) [47]. This aircraft has one class (economy) which consists of 202 passenger seats. The dimensions of the cabin are 36.85 m length × 3.63 m width. The cabin surfaces were divided into small sectors to carry out ray tracing to evaluate the channel impulse response, its delay spread and banwidth. Table 6 introduces the parameter used for cabin communication. Communication is blocked below the surface of the seats.

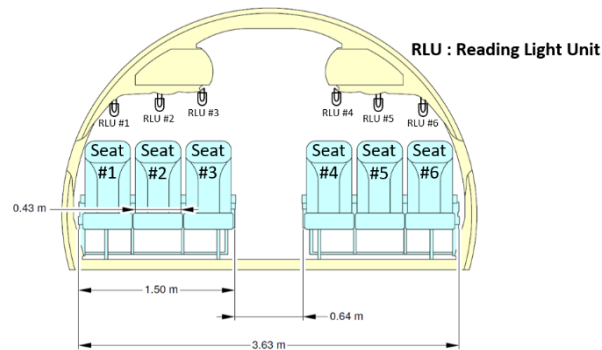

**Fig. 7** Cabin dimentions

**Table 6** The Aircraft Cabin OWC System Parameters

| Parameters | Configurations | |
|---|---|---|
| **Cabin** | | |
| Walls and ceiling reflection coefficient | 0.8 [51] | |
| Floor reflection coefficient | 0.3 [51] | |
| Number of reflections | 1 | 2 |
| Area of reflection element | 5 cm × 5 cm | 20 cm × 20 cm |
| Order of Lambertian pattern, walls, floor and ceiling | 1 [51] | |
| Semi-angle of reflection element at half power | 60º | |
| **Transmitters** | | |
| Number of transmitters' units | 6 | |
| Transmitters locations (RLU) | Above each seat | |
| Number of RYGB LDs per unit | 3 | |
| Transmitted optical power of Red, Yellow, Green and Blue LD | 0.8, 0.5, 0.3 and 0.3 W | |
| Semi-angle at half power | 19º | |
| **Receiver** | | |
| Receiver noise current spectral density | 4.47 pA/√Hz [52] | |
| Receiver bandwidth | 5 GHz | |

### 4.2.2 Aircraft Cabin OWC System Setup and Results

In this part, three devices per passenger have been assumed to examine the resource allocation problem for these devices. The MILP model introduced in Section 3.2 was used here to optimize the resource allocation. The locations of the three devices are assumed as follows: device 1 is placed at the centre of the seat, while devices 2 and 3 are located at a different corners of the seat. The optimized resource allocation of APs and wavelengths is shown in Table 7.

**Table 7** The optimized resource allocation of APs and wavelengths.

| | Passenger #1 | | | Passenger #2 | | | Passenger #3 | | |
|---|---|---|---|---|---|---|---|---|---|
| Device # | Reading Light Unit # | Branch # | Wavelength | Reading Light Unit # | Branch # | Wavelength | Reading Light Unit # | Branch # | Wavelength |
| 1 | 1 | 4 | Red | 2 | 2 | Red | 3 | 3 | Red |
| 2 | 1 | 2 | Green | 2 | 2 | Yellow | 3 | 2 | Green |
| 3 | 1 | 3 | Yellow | 2 | 3 | Green | 3 | 3 | Yellow |
| | Passenger #4 | | | Passenger #5 | | | Passenger #6 | | |
| Device # | Reading Light Unit # | Branch # | Wavelength | Reading Light Unit # | Branch # | Wavelength | Reading Light Unit # | Branch # | Wavelength |
| 1 | 4 | 3 | Red | 5 | 2 | Red | 6 | 4 | Red |
| 2 | 4 | 3 | Yellow | 5 | 3 | Green | 6 | 3 | Yellow |
| 3 | 4 | 2 | Green | 5 | 2 | Yellow | 6 | 2 | Green |

The channel bandwidth, SINR and data rate are shown in Fig. 8. Device number one in all seats, which is located at the centre of the seat has the best channel bandwidth and SINR compared to the other devices. All devices in all seats can support high SINR beyond the threshold and high data rate beyond 7 Gbps.

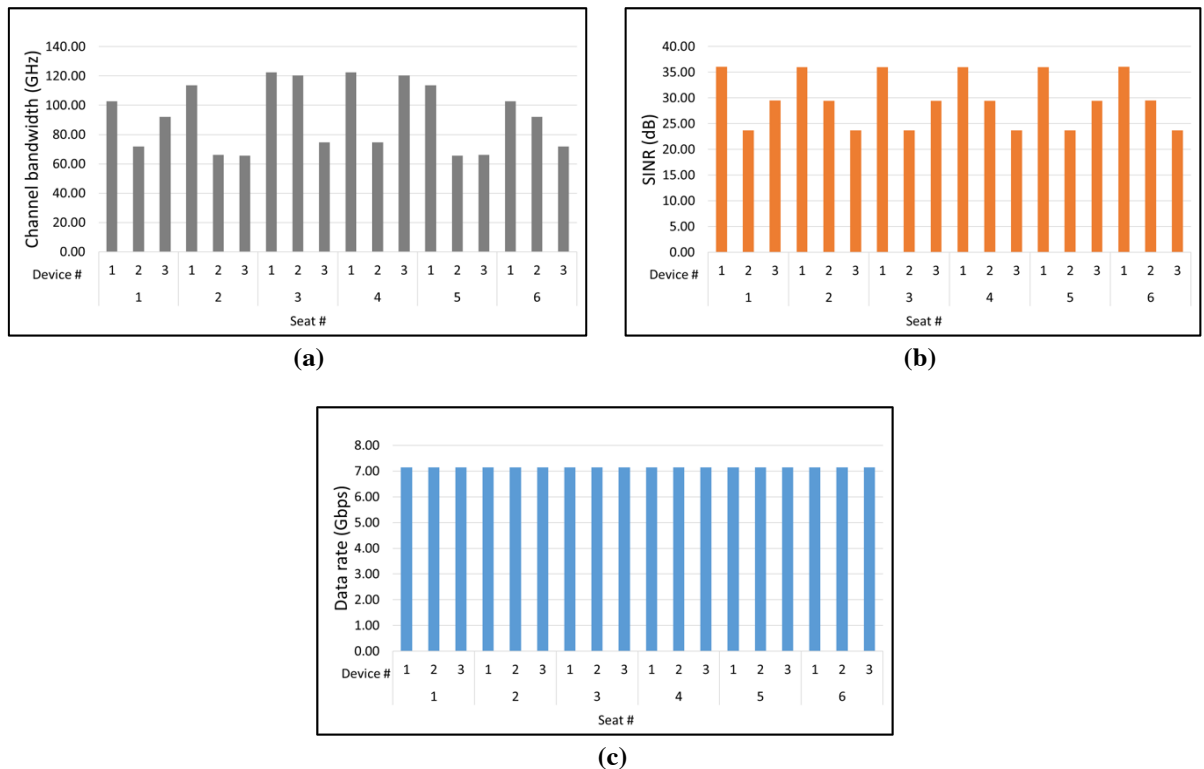

**Fig. 8** (a) Channel Bandwidth, (b) SINR and (c) Data Rate.

### 4.3 Data Centre

This section examines the resource allocation for the downlink OWC system in a data centre. The following sections discuss the data centre configuration and the results.

### 4.3.1 Data Centre OWC System Configuration

The proposed data centre is assumed to be divided into pods, with each pod having dimensions of: 5m length × 6m width × 3m height as shown in Fig. 9. Each pod contains two rows and each row includes 5 racks. In this data centre, each rack has its own switch at the top of the rack. This switch works as a communication coordinator to connect the rack's servers and the racks inside the data centre. The racks' dimensions are shown in Fig. 9. One meter or more has been set between the two rows of racks, as well as between the walls and each row of racks for ventilation. Each rack has its own receiver placed at the centre of the top of the rack to receive signals, as illustrated in Fig. 9. Table 8 introduces the simulation parameters of the data centre.

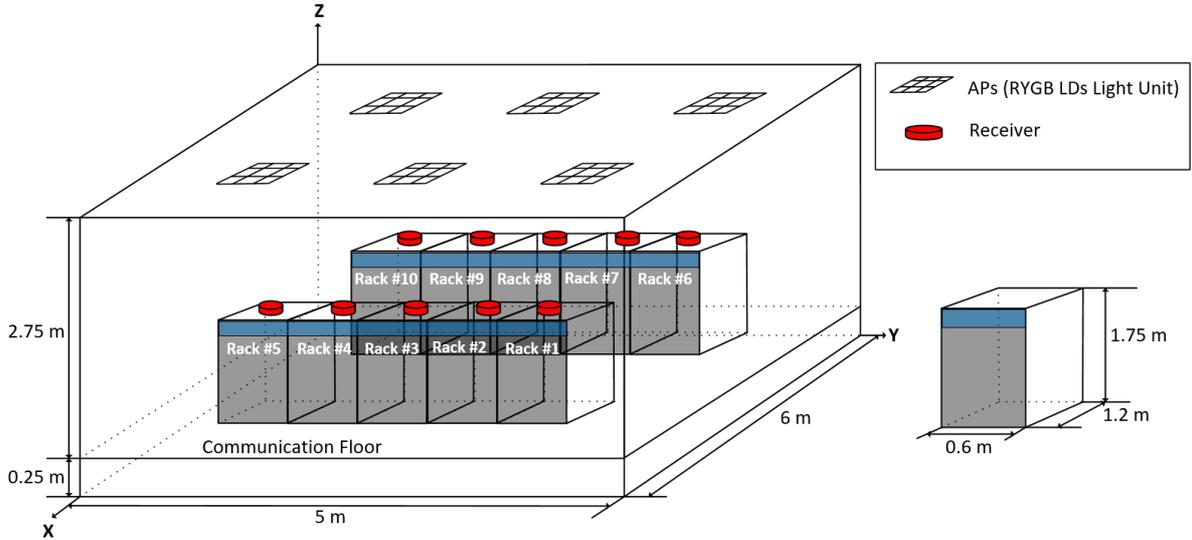

**Fig. 9** Data Center dimensions

**Table 8** The Data Centre OWC Parameters

| Parameters | Configurations | |
| --- | --- | --- |
| **Data Centre** | | |
| Walls and ceiling reflection coefficient | 0.8 [51] | |
| Floor reflection coefficient | 0.3 [51] | |
| Number of reflections | 1 | 2 |
| Area of reflection element | 5 cm × 5 cm | 20 cm × 20 cm |
| Order of Lambertian pattern, walls, floor and ceiling | 1 [51] | |
| Semi-angle of reflection element at half power | 60° | |
| **Transmitters** | | |
| Number of transmitters' units | 6 | |
| Transmitters locations (RLU) | (1.6 m, 1.5 m, 3 m), (1.6 m, 2.5 m, 3 m), (1.6 m, 3.5 m, 3 m) , (4.4 m, 1.5 m, 3 m), (4.4 m, 2.5 m, 3 m) and (4.4 m, 3.5 m, 3 m) | |
| Number of RYGB LDs per unit | 9 | |

| | |
|---|---|
| Transmitted optical power of Red, Yellow, Green and Blue LD | 0.8, 0.5, 0.3 and 0.3 W |
| Semi-angle at half power | 60º |
| **Receiver** | |
| Receiver noise current spectral density | 4.47 pA/√Hz [52] |
| Receiver bandwidth | 5 GHz |

### 4.3.2 Data Centre OWC Setup and Results

A data centre consisting of 10 racks is examined in this section. Each rack has its own receiver located at the centre of the top of each rack. The MILP model in Section 3.2 was used here to assign racks to APs and wavelengths. Table 9 illustrates the optimized resource allocation of APs and wavelengths to each rack.

**Table 9** The optimized resource allocation of APs and wavelengths.

| Rack # | Access Point | Wavelength | Receiver Branch |
|---|---|---|---|
| 1 | 1 | Red | 1 |
| 2 | 1 | Yellow | 4 |
| 3 | 2 | Red | 2 |
| 4 | 3 | Yellow | 2 |
| 5 | 3 | Red | 3 |
| 6 | 4 | Red | 2 |
| 7 | 4 | Yellow | 3 |
| 8 | 5 | Red | 1 |
| 9 | 6 | Yellow | 2 |
| 10 | 6 | Red | 4 |

Fig. 10 illustrates the achieved channel bandwidth, SINR and data rate. All racks support a high channel bandwidth and high SINR above 15.6 dB which is the threshold. Racks assigned to the Red wavelength provide a higher SINR compared to the other racks. All racks can offer a high data rate beyond 7 Gbps.

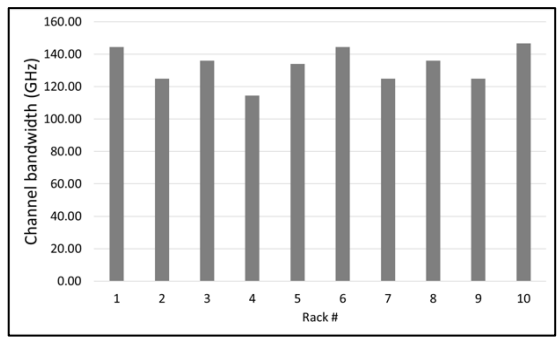
(a)

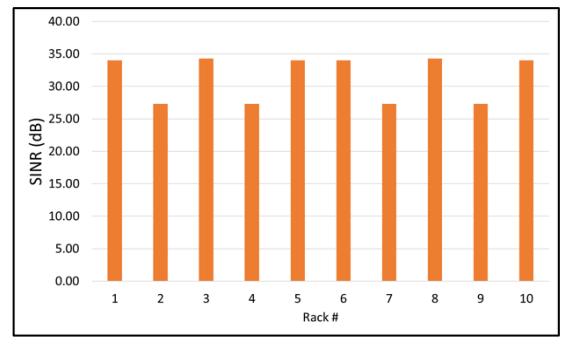
(b)

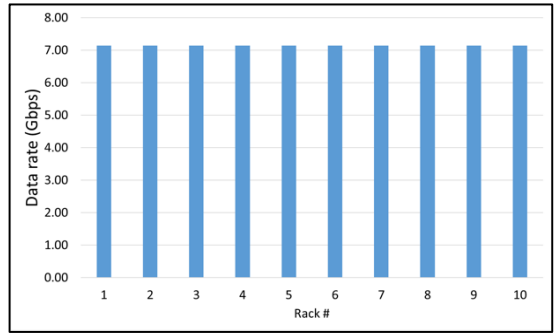

(c)

**Fig. 10 (a)** Channel Bandwidth, **(b)** SINR and **(c)** Data Rate.

## 4. Conclusions

This chapter introduced and discussed the optimization of resource allocation in an optical wireless communication (OWC) system, specifically, a visible light communication (VLC) system. A MILP model was developed to optimize the resource allocation. A wavelength division multiple access (WDMA) system was considered to support multiple users. The optimized resource allocation shows the best assignment of an access point (AP) and wavelength to users. A visible light illumination engine that uses red, green, yellow and blue (RGYB) laser diodes (LD) was utilized to offer high bandwidth for the communication purposes and white light for illumination. An angle diversity receiver (ADR) was utilized to receive signals and reduce noise and interference. Three different indoor environments (office, cabin and data centre) were evaluated using different scenario in this chapter. A scenario of 8 users was examined in the office environment, while, in the cabin, three devices per passenger were evaluated. In the data centre, each rack connects individually to an AP. All scenarios in all environments can support a high data rate beyond 7 Gbps.

## 4. Acknowledgements

The authors would like to acknowledge funding from the Engineering and Physical Sciences Research Council (EPSRC), INTERNET (EP/H040536/1), STAR (EP/K016873/1) and TOWS (EP/S016570/1). All data is provided in the results section of this paper.